\documentclass{article}
\usepackage{spconf,amsmath,graphicx, bbm, amssymb, mathtools, turnstile}

\title{A Random-List Based LAS Algorithm for Near-Optimal Detection in Large-Scale Uplink Multiuser MIMO Systems}
\name{Alexandre A. Pereira Jr and Raimundo Sampaio\_Neto} %\thanks{Thanks to XYZ agency for funding.}}
\address{Centre for Telecommunications Research (CETUC) \\ 
        Pontifical Catholic University of Rio de Janeiro (PUC-Rio) \\
        G\'{a}vea, Rio de Janeiro - Brazil}
\begin{document}
%\ninept
%
\maketitle
\begin{abstract}
Massive Multiple-input Multiple-output (MIMO) systems offer exciting opportunities due to their high spectral efficiencies capabilities. On the other hand, one major issue in these scenarios is the high-complexity detectors of such systems. In this work, we present a low-complexity, near maximum-likelihood (ML) performance achieving detector for the uplink in large MIMO systems with tens to hundreds of antennas at the base station (BS) and similar number of uplink users. The proposed algorithm is derived from the likelihood-ascent search (LAS) algorithm and it is shown to achieve near ML performance as well as to possess excellent complexity attribute. The presented algorithm, termed as random-list based LAS (RLB-LAS), employs several iterative LAS search procedures whose starting-points are in a list generated by random changes in the matched filter detected vector and chooses the best LAS result. Also, a stop criterion was proposed in order to maintain the algorithm's complexity at low levels. Near-ML performance detection is demonstrated by means of Monte Carlo simulations and it is shown that this performance is achieved with complexity of just $O(K^2)$ per symbol, where $K$ denotes the number of single-antenna uplink users.     
\end{abstract}
\begin{keywords}
Massive MIMO, LAS Detection, Near ML Detection
\end{keywords}
\section{Introduction}
\label{sec:intro}

In orther to address the fast growth on the capacity demands on wireless networks experienced in the last years, the use of multiple-input multiple-output (MIMO) technology has been vastly adopted and has become very popular due to the se-veral advantages they offer, including transmit diversity and high data rates \cite{Paulraj03,Jafarkhani05,Tse05}. MIMO techniques are employed at single-user services such as WiFi, WiMAX, and LTE, as well as at multiuser scenarios such as LTE-Advanced and IEEE 802.1 1ac \cite{Liu12,Ghosh10}. Besides, the number of antenna elements employed at these systems has been increasing. The use of large number of antenna elements in MIMO systems has a potential to achieve extremely-high system throughput.  

The multiuser MIMO systems are especially suited when the base station or access point is equipped with lots of antennas. In such scenarios, massive MIMO systems, with an order of 100 antenna elements, that achieve extremely high capacity have been proposed \cite{Marzetta10,Rusek13}. The main issues in realizing such large systems include low-complexity detection and channel estimation. In this work, we approach the complexity of the detection problem in the uplink of a multiuser MIMO system with large number of antenna elements. Well known algorithms such as zero-forcing or minimum mean squared error (MMSE) spatial filtering, sphere decoding \cite{Viterbo99}, and QRM-MLD \cite{Kawai04} require matrix inversion which complexity is proportional to the cubed number of antenna elements. In order to reduce this complexity, several works have been done as detectors based on search strategies such as the likelihood ascent search algorithms \cite{Sun00,Vardhan08, Mohammed08} and Reactive Tabu search \cite{Srinidhi09},  and detection algorithms based on belief propagation \cite{Fukuda13}.

Here, based on the work in \cite{Vardhan08}, we propose a iterative process composed of several one-stage complex LAS detections. The LAS procedure starts with the Matched Filter (MF) detection result and, for each of the following iterations, the starting-vector is derived from random changes of the MF result. The number of iterations is controlled by a stop-criterion that considers the ML cost function of the best LAS result so far as well as the ML cost of an error-free decision. The final decision is the LAS result of the iteration that achieved the least ML cost.

The proposed detector bears some similarities with the work in \cite{Li10}, where the authors proposed the Multiple Initial Vectors (MIV)-LAS and the Multiple Search Candidate Sets (MSCS)-LAS algorithms. As in the proposed RLB-LAS, the MIV-LAS employs a set of different starting-point vectors for the LAS procedure and chooses the best LAS result. The difference between these two algorithms is that the MIV-LAS employs a pre-defined fixed number of starting-point vectors, which can be the result of well-known detection algorithms such as MMSE, ZF and MF, or a completely random gene-rated symbol vector, while the RLB-LAS uses a iterative procedure where on each iteration a new starting-point vector is derived from the MF detection result. As mentioned earlier, the number of iterations/starting-points is controlled by a stop-criterion. The MSCS-LAS uses the MMSE detected result as its starting-point vector and the different LAS results are obtained by different ordering at the symbol changes, which are defined by the search candidate sets. In this work, we choose, as done in \cite{Vardhan08}, a greedy strategy for the ordering of the symbol changes: at each LAS step, the change that provides the greatest decrease in the ML cost is conduced, so there is no pre-defined search candidate set. 

The remainder of this paper is organized as follows. After introducing the system model, the complex LAS criteria and the proposed receiver are presented. Numerical Bit Error Rate (BER) performance simulation results are discussed and the complexity of the proposed detector is analyzed. Finally, some conclusions are made.

\section{System Model}
\label{sec:model}

\begin{figure}
\begin{center}
\includegraphics[ width=8.5cm ,  keepaspectratio=true]{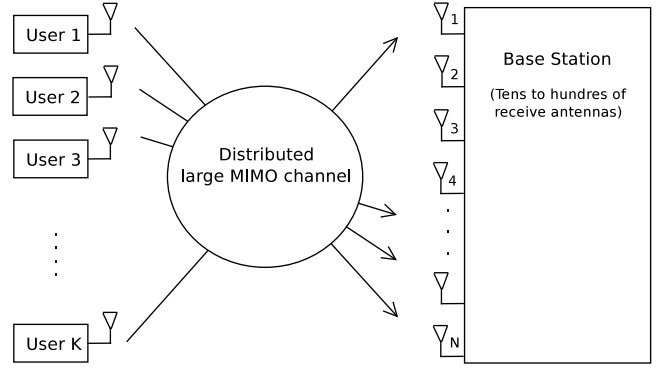}
\caption{Representation of MIMO system.}
\label{mimosys}
\end{center}
\end{figure}

Consider a large-scale MIMO system on the uplink consisting of a base station (BS) with $N$ antennas and $K$ uplink users with one transmit antenna each as depicted in Fig. \ref{mimosys}. All users transmit symbols from a modulation alphabet $\mathbb{B}$. It is assumed that the sampled baseband complex signals are available at the BS receiver.

Let $\mathbf{x} \in \mathbb{C}^{K\times 1}$, $\mathbf{x}=[x_1, x_2, \cdots, x_K]^T$, be the transmitted symbol vector where $x_i$ is the unit energy transmitted complex signal of the $i$-th user. Let $\bf{H} \in \mathbb{C}^{N \times K}$ be the channel gain matrix, such that the $(p,q)$th entry $h_{p,q}$ denotes the complex channel gain from the $q$th user to the $p$th BS receive antenna. In this work, we assume rich scattering conditions so the entries of $\mathbf{H}$ can be modeled as i.i.d. $\mathcal{CN}(0,1)$ aleatory variables, where $\mathcal{CN}(0,1)$ stands for complex gaussian distribution with zero mean and unit variance. Let $\mathbf{y}$ and $\mathbf{n}$ be the received signal vector and the noise vector at the BS, respectively, where the entries of $\mathbf{n}$ are modeled as i.i.d. $\mathcal{CN}(0,\sigma^2_n)$ aleatory variables. The SNR per receiving antenna is defined as $10log_{10}\frac{K}{\sigma^2_n}$, where $\sigma^2_n$ is the noise variance.  The received signal vector can be written as

\begin{equation}
\mathbf{y}=\mathbf{Hx}+\mathbf{n}
\label{eq:receivedsignal}
\end{equation}

The ML solution vector, $\bf{d}_{ML}$, is given by:
\begin{eqnarray}
&\mathbf{d}_{ML}&=\nntstile{\mathbf{d}}{arg min}||\mathbf{y} - \mathbf{Hd}||^2 \nonumber \\
&&=\nntstile{\mathbf{d}}{arg min}\mathbf{d}^H\mathbf{H}^H\mathbf{H}\mathbf{d} - 2Real\{\mathbf{y}^H\mathbf{Hd}\}.
\label{eq:MLcost}
\end{eqnarray}
whose complexity is exponential in $K$ and $||\mathbf{x}||$ denotes the Euclidean norm of the vector $\mathbf{x}$. $(\cdot)^H$ stands for complex Hermitian conjugate. A low-complexity high-performance detection algorithm is presented next.

\section{Proposed Receiver}
\label{sec:proprec}
\subsection{Complex LAS Procedure}
\label{sec:CLAS}
In this section we present a complex valued alternative of the LAS algorithm proposed in \cite{Vardhan08}. This algorithm consists in a series of likelihood-ascent search stages where in each stage a sequence of one-symbol update is conduced such that the likelihood monotonically increases from one iteration to another. Once a local minimum is reached, one could pass to a second stage where two-symbol updates are conduced, them 3-symbol updates and so forth. In order to maintain a low-complexity algorithm, in this work, we consider only one-symbol updates.

Starting from an initial solution $\mathbf{d}^{(0)}$, the algorithm searches for a new solution, $\mathbf{d}^{(1)}$, that differs from $\mathbf{d}^{(0)}$ in exactly one position (one-symbol update), such that the ML cost of the new solution is lesser than the ML cost of the current one. The algorithm continues this procedure until a local minimum is achieved, i.e., there is no new solution that differs from the current solution by one symbol and that has a lower ML cost. The ML cost function after the $k$th iteration is given by
\begin{equation}
C^{(k)}={\mathbf{d}^{(k)}}^H\mathbf{H}^H\mathbf{H}\mathbf{d}^{(k)} - 2Real\{\mathbf{y}^H\mathbf{Hd}^{(k)}\}.
\label{eq:MLcost}
\end{equation}

In order to guarantee a monotonic decrease in the ML cost, lets consider the following development. Assuming that the $p$th symbol is updated in the $(k+1)$th iteration, $p=1, 2, \cdots, K$, the update rule can be written as
\begin{equation}
\mathbf{d}^{(k+1)}=\mathbf{d}^{(k)}+\lambda_p^{(k)}\mathbf{e}_p,
\label{eq:updaterule}
\end{equation}
where $\mathbf{e}_p$ denotes the unit vector where its $p$th entry is one and all the other entries are zero. The possible values of $\lambda_p^{(k)}=d_p^{(k+1)}-d_p^{(k)}$ depends on the constellation being employed and the current value of the $p$th entry of the vector $\bf{d}^{(k)}$. For example, for 4-QAM constellation such as $[1+j, 1-j, -1-j, -1+j]$, if the $p$th entry of the vector $\bf{d}^{(k)}$ is $d_p^{(k)}=1+j$, the possible values of $\lambda_p^{(k)}$ are $[-2, -2j, -2-2j]$. Using ($\ref{eq:MLcost}$) and ($\ref{eq:updaterule}$), and defining $\bf{G}$ as
\begin{equation}
\mathbf{G} \triangleq \mathbf{H}^H\mathbf{H},
\label{eq:defG}
\end{equation}
the cost difference, $\Delta C_p^{k+1}=C_p^{k+1}-C_p^k$, can be written as
\begin{equation}
\Delta C_p^{k+1}(\lambda_p^{(k)})=|\lambda_p^{(k)}|^2(\mathbf{G})_{p,p} -2Real\{ {\lambda_p^{(k)}}^* z_p^{(k)} \},
\label{eq:deltacusto}
\end{equation}
where ($\mathbf{G})_{p,p}$ denotes the $p$th entry of the main diagonal of $\mathbf{G}$, $\mathbf{z}^{(k)}=\mathbf{H}^H(\mathbf{y}-\mathbf{Hd}^{(k)})$, and $z_p^{(k)}$ is the $p$th entry of the $\mathbf{z}^{(k)}$ vector. In each iteration, the value of $\lambda_p$ is evaluated as 
\begin{equation}
\lambda_p=\nntstile{\lambda \in \mathbb{P}}{arg min}\Delta C_p(\lambda),
\label{eq:lambdap}
\end{equation}  
where $\mathbb{P}$ is the set of all possible $\lambda_p$ values. To chose the position, $s$, that will be actually updated at the current iteration, the algorithm considers the position that gives the greater decrease in the cost function, i. e. 
\begin{equation}
s=\nntstile{p \in [1, \cdots, N]}{arg min}\Delta C_p,
\label{eq:s}
\end{equation}  
If at a given iteration $\Delta C_s$ is not negative, then a local minimum was reached and the algorithm stops.

 \subsection{Proposed Random-List Based LAS Detector}
\label{sec:RLB-LAS}
The key idea behind the proposed RLB-LAS algorithm is to avoid local minima by conducing several iterative complex LAS procedures. In order to maintain the algorithm's complexity at low level, at the first iteration, the MF symbol vector result, $\bf{d}^{(0)}_{MF}$, is employed as the starting point vector of the LAS procedure and the LAS result, $\mathbf{d}_{MF}$, is elected as the current decision. For each one of the following iterations, steps $1$ to $7$ are conduced:

% At the end of each LAS procedure, the ML cost of the resulting symbol vector is compared with the  employing several different LAS starting point vectors, and, for each starting vector, the complex LAS procedure is conduced and the result that achieves the least ML cost is elected as the detected symbol vector. The starting point vector set is composed by the MF detector result, $\bf{d}^{(0)}_{MF}$, and by $Np-1$ other symbol vectors, where $Np$ represents the number of elements in $\mathbb{V}$. Each one of the $Np-1$ starting point symbol vectors are obtained from the MF result as follows:

1) Set $\mathbf{d}^{(0)}_{m}=\mathbf{d}^{(0)}_{MF}$, where $m$ stands for the current iteration;

2) Chose $c$ randomly from a uniform distribution over $[1, \cdots, K]$, where $c$ represents the number of symbols to be changed from the MF result;

3) Select $c$ values sampled uniformly at random, without replacement, from the integers $[1, \cdots, K]$, to form the indices, $\bf{i}=[i_1, \cdots, i_c]$, of the entries of the MF result vector to be changed;

4) For $l=1, 2, \cdots, c$, select $\mathbf{d}^{(0)}_{m}(i_l)$ from the symbol constellation $\mathbb{B}$ sampled uniformly at random. Steps $1$ trough $4$ defines a new starting point symbol vector for the complex LAS procedure;

5) Perform the complex LAS procedure with $\mathbf{d}^{(0)}_{m}$ as star-ting-point;

6) If the ML cost of the LAS result symbol vector is less then the ML cost of the current decision, update the current decision as the LAS result of the current iteration;

7) Go to next iteration.  

We name the proposed detector MF-RLB-LAS because all the starting point-symbol vectors employed are derived from the MF result by a random number of aleatory symbol changes. To complete the algorithm, a stop criterion is needed. A simple strategy would be to determine a fixed number of iterations, $Np$. This strategy may not be appropriate, causing a performance loss, especially at high signal-to-noise ratio (SNR). To address this issue, the number of iterations should depend on the quality of the ML cost of the LAS result, $\mathbf{d}_{MF}$. If the quality is poor, then a large number of iterations is needed in order to avoid local minima. On the other hand, if the $\mathbf{d}_{MF}$ ML cost quality is already good, then a small number of iterations is preferred. For this purpose, we use the quality metric employed in \cite{Datta12}. It is determined in terms of the closeness of the ML cost of a given detected vector to a value obtained using the statistics of the ML cost for the case when the received vector is detected error-free. Note that the ML cost of a error-free detection corresponds to $||\bf{n}||^2$, which is Chi-squared distributed with $2N$ degrees of freedom with mean $N\sigma^2$ and variance $N\sigma^4$. In \cite{Datta12}, the quality metric is defined as the difference between the ML cost of the detected vector and the mean of $||\mathbf{n}||^2$, scaled by the standard deviation, i.e., the quality metric of $\mathbf{d}$ is defined as 

\begin{equation}
\phi(\mathbf{d})=\frac{|| \mathbf{y}-\mathbf{H}\mathbf{d}||^2-N\sigma^2}{\sqrt{N}\sigma^2}
\label{eq:metric}
\end{equation}  

The metric at (\ref{eq:metric}) is referred as the \textit{standardized ML cost} of $\mathbf{d}$. A small value of $\phi(\mathbf{d})$ represents an increased closeness to the ML solution. Therefore, it is desired to chose $Np$ as a increasing function of $\phi(\mathbf{d})$. An exponential function is chosen, and an minimum value, $Np_{min}$, is also employed, i.e.
\begin{equation}
Np=\lceil max( c_1exp(\phi(\mathbf{d})),Np_{min} ) \rceil
\label{eq:Np}
\end{equation}  
where $c_1$ is a metric parameter and $\lceil a \rceil$ stands for the least integer greater than a.

One last observation concerning $Np$ is that, due to the low effectiveness of the MF receiver to deal with the interfe-rence problem the $Np$ value resulting from (\ref{eq:Np}) for the LAS result $\mathbf{d}_{MF}$ may become unnecessarily large for high SNR. This large number of iterations may lead to an unnecessary increase on the algorithm complexity. To avoid this problem, at each current decision update, we also update $Np$ using the \textit{standardized ML cost} of the current decision symbol vector. At the end of each iteration, if the iteration number is greater then $Np$, the algorithm stops and the current decision signal vector is elected as the final detected signal vector.

\section{Numerical Results and System Complexity}
\label{sec:numres}
\subsection{RLB-LAS Performance}
\begin{figure}
\begin{center}
\includegraphics[ width=9cm ,  keepaspectratio=true]{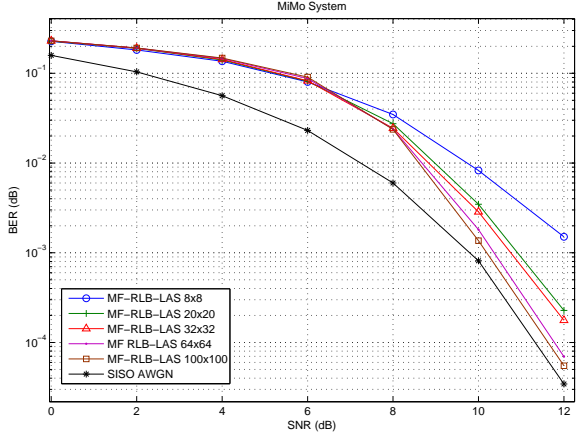}
\caption{BER performance of the RLB-LAS detector.}
\label{result1}
\end{center}
\end{figure}

The Bit Error Rate (BER) performance of the proposed detectors is analyzed in this section. These BER curves were eva-luated by Monte Carlo simulations of a 4-QAM modulation. The results are an average of 1000 independent simulation runs with 10 symbol vector transmissions per run over a rich scattering Rayleigh fading model channel. The following parameters are used in the simulation: $Np_{min}=2$, $c_1$=5. Fig. \ref{result1} shows BER results of the MF-RLB-LAS detector for several values of $K=N$, as well as the AWGN-only Single-input Single-output (SISO) result for comparison. Perfect channel knowledge is assumed at the receivers. From Fig. \ref{result1}, as in  others LAS algorithms \cite{Vardhan08}, it can be observed that the performance of the proposed MF-RLB-LAS improves with increa-sing $K=N$, getting close to the AWGN-only SISO channel result. It can be noticed that for $K=N=20$ it achieves an uncoded BER of $10^{-3}$ at almost 1 dB away from the AWGN-only SISO performance. Increasing $N=K$ this gap gets even smaller. 

\begin{figure}[t]
\begin{center}
\includegraphics[ width=9cm ,  keepaspectratio=true]{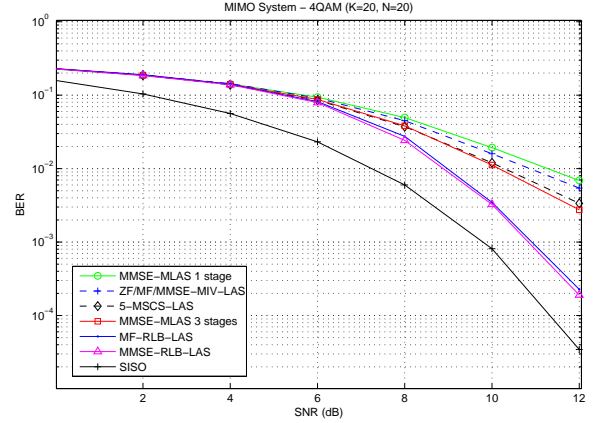}
\caption{BER performance of several MIMO uplink detectors.}
\label{result1.2}
\end{center}
\end{figure}

\begin{figure}
\begin{center}
\includegraphics[ width=9cm ,  keepaspectratio=true]{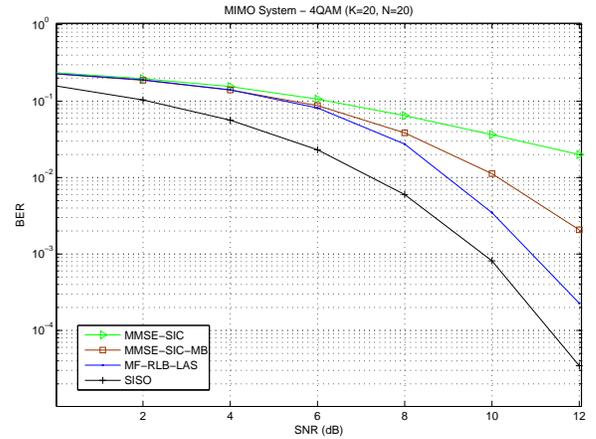}
\caption{BER performance of the MF-RLB-LAS, MMSE-SIC and MMSE-SIC-MB MIMO uplink detectors.}
\label{result1.3}
\end{center}
\end{figure}

This result illustrates the ability of the LAS algorithm family to approach single-antenna AWGN performance even in a large MIMO scenario, removing spatial interference from other antennas. The advantage of the proposed MF-RLB-LAS algorithm over other LAS algorithms proposed in \cite{Vardhan08} and \cite{Li10} is that the MF-RLB-LAS approaches AWGN SISO performance as the number of antenna elements increases faster, as can be seen in Fig. \ref{result1.2}, where the BER results of the proposed MF-RLB-LAS algorithm are shown as well as the BER results of the MMSE-MLAS with one and three stages, the results of the MIV-LAS, MSCS-LAS and an alternative RLB-LAS that employs the MMSE filter detection result instead of the MF result as $\mathbf{d}^{(0)}$, termed MMSE-RLB-LAS. Also, in Fig. \ref{result1.3} is shown, besides the MF-RLB-LAS results, the performance curves of two other detectors that employ  successive interference cancellation (SIC) and multi-branch (MB) technique, the MMSE-SIC and MMSE-SIC-MB \cite{Patel94, Lamare11, Leonel14}. The curves presented in Fig. \ref{result1.3} were separated from Fig. \ref{result1.2} for clarity. It can be noticed that the RLB-LAS detectors achieved the best results and, besides, the performance curves of the MF-RLB-LAS and the more complex MMSE-RLB-LAS alternative were very similar. Also, as shown in the next subsection, the MF-RLB-LAS requires less floating point operations then all the other detectors.

\subsection{Complexity analisys}

\begin{figure}[t]
\begin{center}
\includegraphics[ width=9cm ,  keepaspectratio=true]{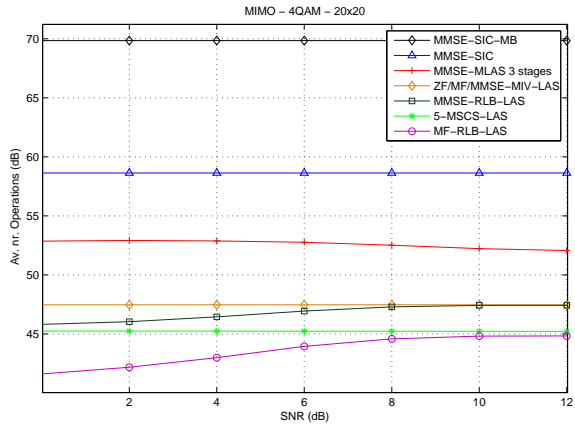}
\caption{Complexity results of the MF-RLB-LAS, MMSE-RLB-LAS, 3-stage MMSE-MLAS, MMSE-SIC and MMSE-SIC-MB algorithms in average number of floating-point operations per symbol as function of SNR for 20x20 MIMO system with $4$-QAM.}
\label{result2.2}
\end{center}
\end{figure}

Figure \ref{result2.2} shows the average number of floating-point ope-rations (flops) versus SNR for a MIMO system with $K=N=20$ for the MF-RLB-LAS, MMSE-RLB-LAS, MMSE-SIC, MMSE-SIC-MB, ZF/MF/MMSE-MIV-LAS, 5-MSCS-LAS and 3 stage MMSE-MLAS algorithms. The number of flops were computed using the Lightspeed Matlab toolbox \cite{lightspeed07}.  

\begin{figure}
\begin{center}
\includegraphics[ width=9.5cm ,  keepaspectratio=true]{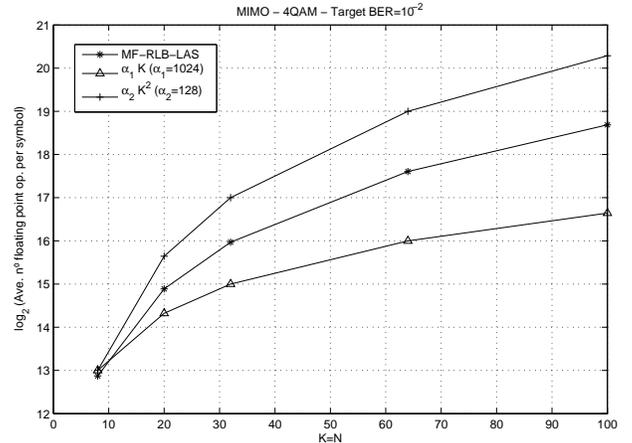}
\caption{Complexity of the MF-RLB-LAS algorithm in average number of floating-point operations per symbol as function of $K=N$ with $4$-QAM.}
\label{result2}
\end{center}
\end{figure}

While ML decoder gets exponentially complex in $K$, the MF-RLB-LAS complexity (in average number of flops per symbol) is only $\mathcal{O}(K^2)$ as can be seen in Fig. \ref{result2}, that ilustrates the complexity result as a function of $K=N$ for a target BER of $10^{-2}$ of the MF-RLAS detector and two other bounding curves. Table \ref{table1} shows the average number of flops per symbol of the proposed MF-RLB-LAS detector as well as the average number of operations of the MMSE-MLAS and of the ML detector for comparison. As can be noticed, the proposed MF-RLB-LAS detector achieves similar performance results with less operations.
  
\begin{table}
 \centering
\small
\begin{tabular}{| c | c | c | c |}
\hline

 $K=N$ & MF-RLB-LAS & \begin{tabular}{c} MMSE-MLAS \\ (3-stage) \end{tabular} & ML \\ \hline
$8X8$         & $ 7.5 \times 10^{3}$ & $2.2 \times 10^4$ & $6.9 \times 10 ^{7}$ \\ \hline
$20X20$     &  $3 \times 10^{4}$    & $1.7 \times 10^{5}$ & $1.1 \times 10^{12} $ \\ \hline
$100X100$ &  $4.2 \times 10^{5}$ &$6.6 \times 10^6$  & $1.0 \times 10^{64}$ \\ \hline

\hline
\end{tabular}
\caption{Average number of floating-point operations for target BER of $10^{-2}$.}
\label{table1}
\end{table}

\section{Conclusions}
This work presented a LAS MIMO uplink detector that achieves near-ML performance with just $\mathcal{O}(K^2)$ complexity. It randomly creates several starting-point candidate vectors from the Matched Filter detected vector and, through an iterative sequence of LAS procedures, the MF-RLB-LAS algorithm chooses the LAS result with the best ML cost as its final decision. Monte Carlo simulations were performed in order to show that this algorithm can achieve BER results close to the SISO AWGN-only channel. The MF-RLB-LAS results were compared to other LAS algorithm, including the $3$-stage MMSE-MLAS algorithm, the MIV-LAS, the MSCS-LAS, and an alternative version of the RLB-LAS, the MMSE-RLB-LAS, and other MIMO detectors such as the MMSE-SIC and MMSE-SIC-MB. It was shown that the proposed algorithm achieved better BER and complexity results then the MMSE-MLAS, MMSE-SIC and MMSE-SIC-MB, and better complexity results then the MMSE-RLB-LAS with similar performance.

\label{sec:print}

\bibliographystyle{IEEEbib}
\bibliography{RLAS}

\end{document}